\documentclass{PoS}

\title{Review on Active Galactic Nuclei at hard X-ray energies}

\ShortTitle{AGN at hard X-ray energies}

\author{\speaker{L. Bassani}$^{a}$, M. Molina$^{a}$, A. Malizia$^{a}$,
F. Panessa$^{b}$, R. Landi$^{a}$, A. Bazzano$^{b}$, P. Ubertini$^{b}$,
A. J. Bird$^{c}$ and J. B. Stephen$^{a}$\\
\llap{$^{a}$} INAF/IASF Bologna, Italy\\
\llap{$^{b}$} INAF/IAPS Rome, Italy\\
\llap{$^{c}$} School of Physics and Astronomy, University of Southampton, UK\\
E-mail: \email{bassani@iasfbo.inaf.it}}                    
        
\abstract{Hard X-ray surveys are an important tool for the study of active galactic nuclei (AGN): 
they provide almost an unbiased view of absorption in the extragalactic population, allow 
the study of spectral features such as reflection and high energy cut-off which would otherwise be 
unexplored and favour the discovery of some blazars at high redshift.
Here, we present the absorption properties of a large sample of INTEGRAL detected AGN, including an 
update on the fraction of Compton thick objects. For a sub-sample of 87 sources, which represent a 
complete set of bright AGN, we will discuss the hard X-ray (20--100\,keV) spectral properties,
also in conjunction with Swift/BAT 58 month data, providing 
information on BAT/IBIS cross-calibration constant,
average spectral shape and spectral complexity. For this complete sample, we will also 
present broad-band data using soft X-ray observations, in order to explore the complexity of AGN 
spectra both at low and high energies and to highlight the variety of shapes. 
Future prospects for AGN studies with INTEGRAL will also be outlined.}

\FullConference{An INTEGRAL view of the high-energy sky (the first 10 years) - 
9th INTEGRAL Workshop and celebration of the 10th anniversary of the launch,\\
		October 15-19, 2012\\
		Bibliotheque Nationale de France, Paris, France}

\begin{document}

\section{Introduction}
The study of active galactic nuclei (AGN) above 10\,keV is essential if one 
wants to study non-thermal processes and observe those sources
which are strongly affected by absorption in the soft X-ray band. 
Another advantage is the possibility of having information on spectral features such as the high energy cut-off
and the reflection fraction, which cannot be explored with observations performed below 10\,keV.
The determination of these parameters is important for many reasons: they  
provide an insight into the physical properties
of the region around the central power source, play a key role in synthesis models  
of the cosmic X-ray background and are important ingredients 
for unification theories and torus studies (e.g \cite{urry95, elitzur06, elitzur12}).

\section{The sample}

\begin{figure}
\centering
\includegraphics[width=0.4\linewidth]{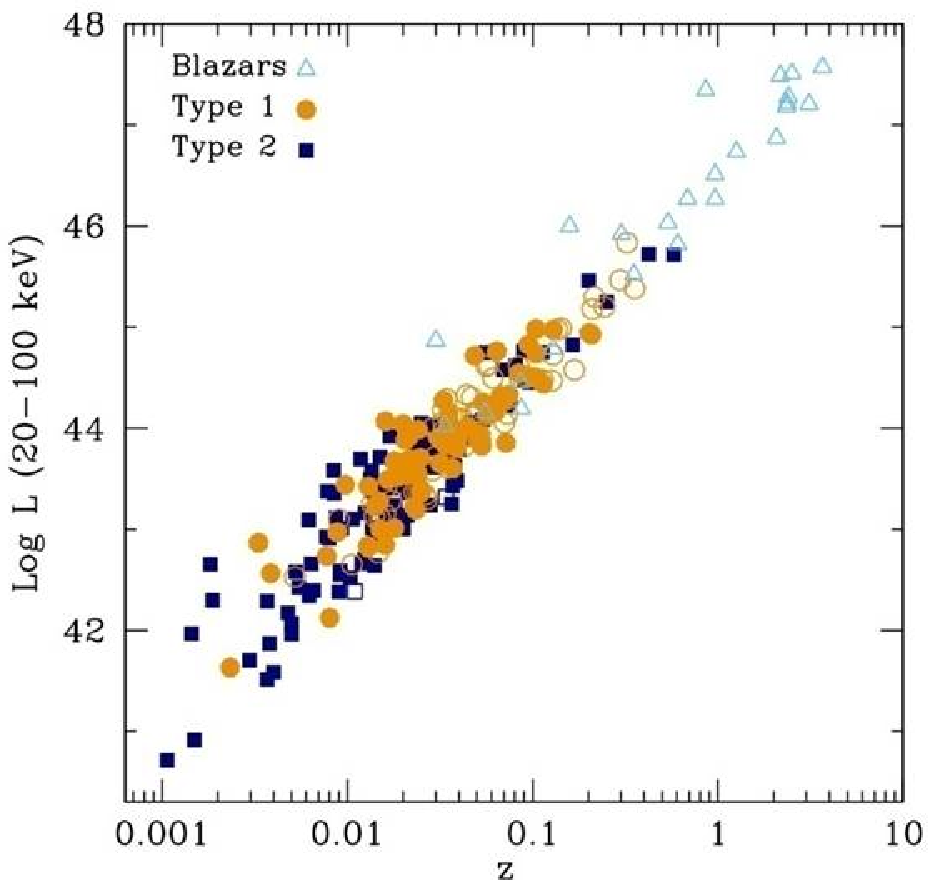}
\includegraphics[width=0.4\linewidth]{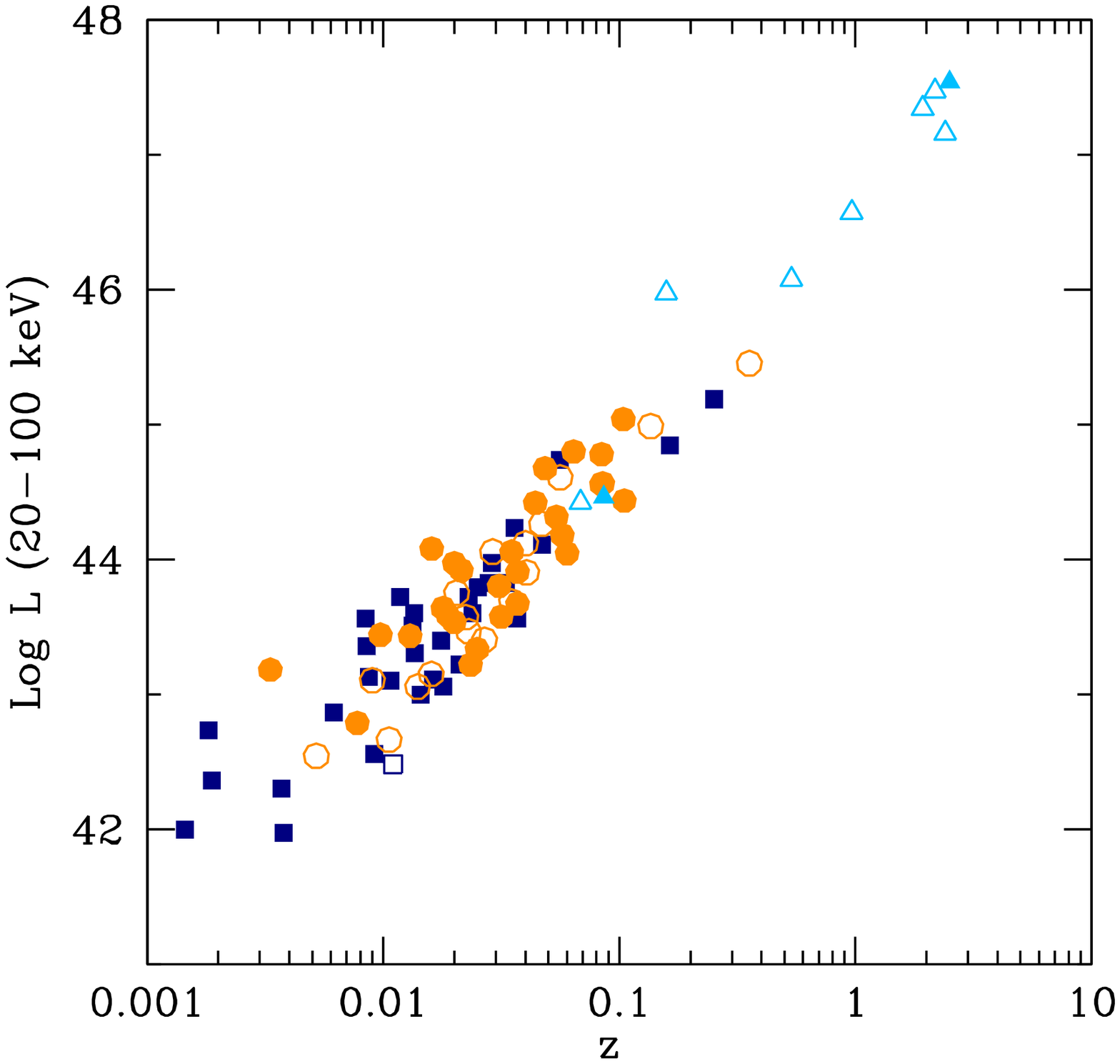}\\
\caption{\small{\emph{Left Panel}: Hard X-ray luminosity {\it vs} redshift for the 
INTEGRAL sample of 272 AGN. Circles are type 1 AGN, squares
are type 2 AGN and triangles are blazars. Filled symbols represent 
objects with absorption measured while   open symbols refer to sources with upper limits N$_{\rm H}$.
 \emph{Right Panel}: Same as in the left panel but using only objects belonging to the complete AGN sample.}}
\label{z_lum}
\end{figure}

In the 4$^{\rm th}$ survey \cite{bird10} made by INTEGRAL/IBIS \cite{winkler03}
there are 234  objects which have been identified 
with AGN. To this set of objects, one can add 38 galaxies listed in the INTEGRAL all-sky survey by 
\cite{krivonos07} updated on the website\footnote{http://hea.iki.rssi.ru/integral/survey/catalog.php}
but not included in the Bird et al. catalogue due to the different sky coverage. 
The final dataset comprises 272 AGN (last update March 2011), which represents the most complete view 
of the INTEGRAL extragalactic sky to date \cite{malizia12}.
This sample has two great strengths: a) all sources have optical spectra, which means a secure identification 
and a measured redshift (the only exception is the BL Lac object RX J0137.7+5814 for which no redshift is 
available), and b) all sources have X-ray data available, which provides a measure of the source intrinsic 
absorption. In the left panel of figure \ref{z_lum}, the whole sample is reported in the classical 
20--100\,keV luminosity {\it vs} redshift plot. The luminosities have been calculated for all sources 
assuming H$_{0}$=71\,km\,s$^{-1}$\,Mpc$^{-1}$ and q$_{0}$ = 0. From this figure, it can be estimated 
that our sensitivity limit is around 1.5$\times$10$^{-11}$erg\,cm$^{-2}$\,s$^{-1}$. 
We find that the source redshifts span from 0.0014 to 3.7 with a mean at z=0.1477,
while the Log of 20--100\,keV luminosities ranges from 40.7 to $\sim$48 with a mean at around 
10$^{46}$ erg\,s$^{-1}$. NGC 4395 (a Seyfert 2) is the closest and least luminous AGN seen by INTEGRAL, 
while IGR J22517+2218 (a broad line QSO) is the farthest and most luminous;
the former hosts a relatively small central black hole (M$\sim$10$^{4}$-10$^5$\,M$_{\odot}$), 
while the latter houses a more massive one (M=10$^9$ M$_{\odot}$). Thus the INTEGRAL AGN catalogue spans 
many order of magnitudes in source luminosity and black hole mass, indicating that it covers a large range 
in Eddington luminosities and accretion rates.
 
Although useful for statistical and population studies, this catalogue is too large and contains too
many weak sources to be appropriate for hard X-ray spectral studies 
(either individually or in combination with low energy observations). For this purpose we have created 
a subsample of objects which constitute a complete sample of AGN selected in the 20--40\,keV band.
Full details on the extraction of this complete sample are given in \cite{malizia09}. It consists of 87 
active galaxies of various optical classification: 41 type 1 AGN (Seyfert 1, 1.2 and 1.5), 33 type 2 AGN (Seyfert 1.9
and 2), 5 narrow line Seyfert 1s (NLSy1s) and 8 Blazars (QSOs and BL Lacs). In the right panel of Figure \ref{z_lum},
the objects in this subsample are displayed in the 20--100\,keV luminosity {\it vs} redshift plot;
it is clear from a comparison between the two  panels that the data sets are similar in parameter distribution 
so that we can safely assume that the complete sample is representative of the entire population of AGN selected 
in the hard X-ray band.

\section{Absorption properties}
In the following, N$_{\rm H}$=10$^{22}$ cm$^{-2}$ is assumed as the dividing line between
absorbed and unabsorbed sources: this is the value conventionally used because it corresponds to a column 
density sufficiently high to hide an AGN broad line region (BLR, \cite{silverman05}). We also consider as 
Compton thick an object with a column density in excess of 1.5$\times$10$^{24}$\,cm$^{-2}$. It is worth noting 
that for a number of objects we did not measure any absorption in excess to the Galactic value and have
therefore used the Galactic column density as an upper limit to the source intrinsic absorption.
With the assumptions made above, the fraction of absorbed objects present in our sample is 49\%,  
while that of Compton thick objects is around $\sim$6\%, in full agreement with previous estimates 
available in the literature (\cite{malizia09}, \cite{burlon11}).

Despite the fact that hard X-ray instruments are the least biased in terms of detecting absorbed AGN, they
still miss some Compton thick objects, essentially those with weak (intrinsic) fluxes and at large distances. 
This has been fully discussed by \cite{malizia09} and \cite{burlon11}, who have shown that once the
correction for this bias is applied, the real intrinsic fraction of Compton-thick AGN 
is around 20-24\%. In particular, in our previous work (\cite{malizia09}) we have adopted a redshift cut 
(z=0.015 or 60\,Mpc) in our complete sample of INTEGRAL AGN in order to remove the bias and
to probe, although only locally, the entire AGN population.
Following this reasoning and using the entire AGN sample,
we should be able to expand our previous analysis and to confirm our initial hypothesis, 
having in mind that the present sample although enlarged is not complete. To allow a comparison with the work 
of \cite{malizia09}, we have divided the sample in the same 
redshift bins (up to z=0.57 considering only those AGN with Log L$_{\rm 20-100keV}\leq$46) 
and plotted  the fraction of absorbed (N$_{\rm H}>$10$^{22}$ cm$^{-2}$) objects compared 
to the total number of AGN in these bins. The result is shown in Figure \ref{2_xx} 
(left panel), for the enlarged  (red points) and  complete sample (black points). A number of considerations can be made from this figure.
First, the bias in z is still present as we keep observing a trend of decreasing
fraction of absorbed objects as the redshift increases. Second, in the first bin the fraction of 
absorbed objects remains the same as found before: 
in particular we find that, out of 66 objects, 53 (or 80\%) have a column density $\geq$10$^{22}$ and 11 (or 
17\%$\pm$3\%) are Compton thick. Taking into consideration the fact that the entire sample is not complete, 
these results are in close agreement with those found previously and confirm 
the original suggestion that our survey is able to pick up all AGN, even the most absorbed ones, but only in 
the local Universe.
Finally, the fraction of absorbed (and Compton thick) AGN has increased in the second bin from 
35\% to 57\% (including 4 Compton thick sources), implying that as the INTEGRAL survey progresses, we are able 
to pick up more absorbed objects among those which are  
distant and faint and therefore lost in previous catalogues.

We thus conclude that the bias affecting deep hard X-ray surveys of AGN is real but 
negligible if we deal with objects located 
in the nearby Universe, where the {\textquotedblleft{true}\textquotedblright} fraction of absorbed  
and Compton thick objects can be estimated with 
some precision at 80\% and $\sim$17\% respectively. Similar indications are now emerging also from X-ray 
observations of AGN samples selected in other wavebands. For example \cite{brightman11} 
studied the AGN present in the extended IRAS 12 micron sample \cite{rush93} for which
an XMM-Newton observation is available and found that the fractions of obscured and Compton thick objects  
are 62$\pm$5\% and 20$\pm$4\% respectively. 
\cite{akylas09} presented instead the XMM-Newton spectral analysis of all
38 Seyfert galaxies in the Palomar spectroscopic sample of \cite{ho97}: they estimate the fraction of 
absorbed nuclei to be 75$\pm$19\%,  while that of Compton-thick sources to be 15-20\%. So the numbers obtained 
from samples selected in various wavebands seem to converge at least for what concern the 
local Universe, while at high z the situation is by far less certain.

\section{Hard X-ray spectra}

\begin{figure}
\centering
\includegraphics[width=0.4\linewidth]{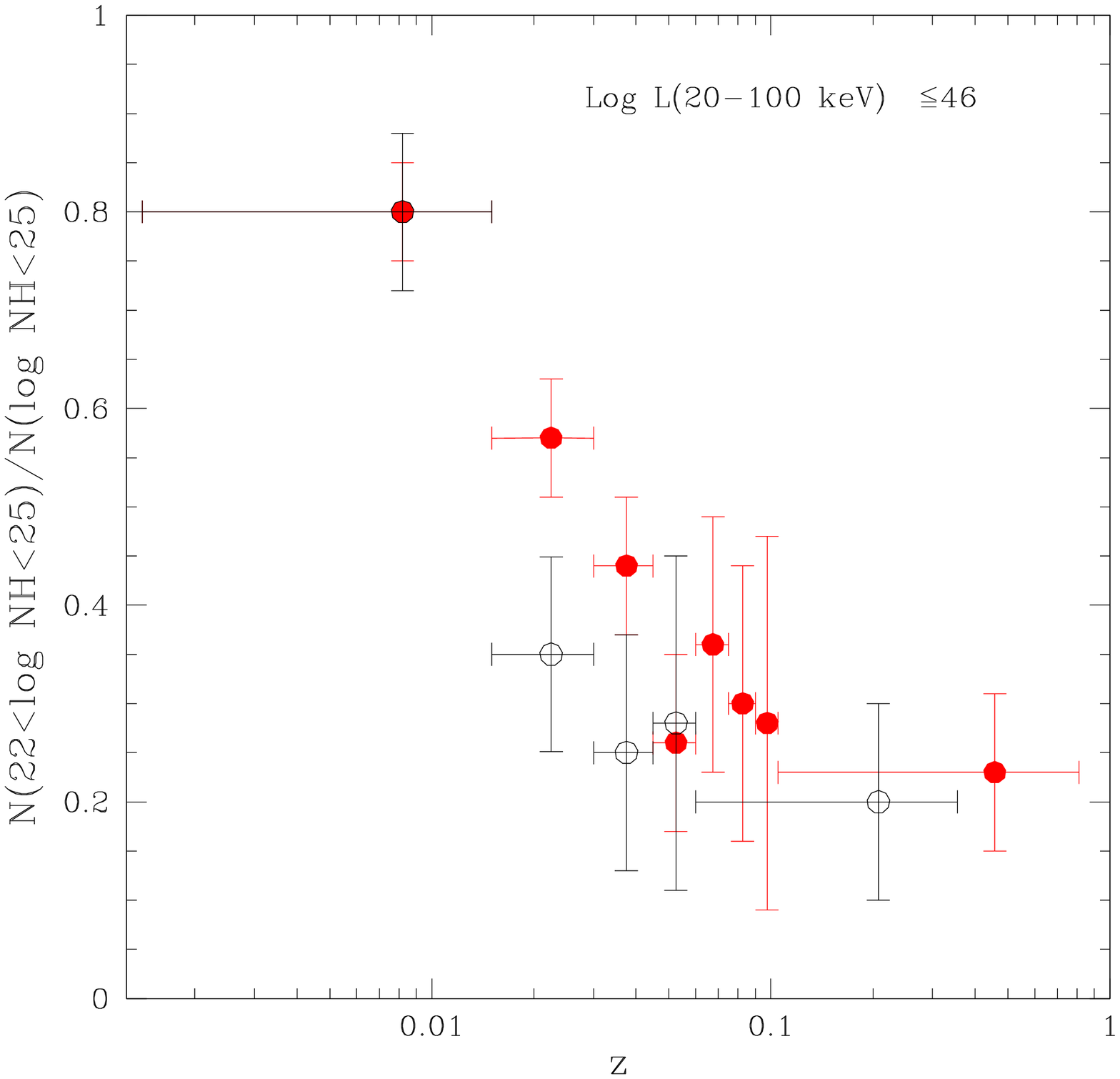}
\includegraphics[width=0.3\linewidth]{3C273_cont.eps}\\
\caption{\small{\emph{Left Panel}: Fraction of absorbed objects compared to the total
number of AGN as a function of redshift from enlarged  (red
points,\cite{malizia12}) and complete (black points,\cite{malizia09}) samples respectively. 
\emph{Right Panel:} \texttt{pexrav} model applied to the combined IBIS/BAT data of 3C 273 with the high energy cut-off fixed to 1\,MeV.}}
\label{2_xx}
\end{figure}

One problem effecting the study of the hard X-ray spectra of AGN is the low statistical significance of the data
which often limits the measurements of parameters such as the reflection fraction and the high energy cut-off.
A way to overcome this problem is to combine data from INTEGRAL/IBIS and Swift/BAT. Before doing so, it is important 
to cross calibrate the two instruments: one way is the classical method of using the Crab \cite{ki05} , another is to use
average spectra of large samples of objects belonging to the same class. 
About 80 objects of the 87 listed in the complete INTEGRAL sample have also BAT data available from the 58-month 
survey \footnote{http:http://heasarc.gsfc.nasa.gov/docs/swift/results/bs58mon/}. A simple power law
fit to the combined IBIS/BAT points of all these AGN provides 
a cross-calibration constant of 1.14$\pm$0.04, sufficiently close to one to justify the joint use of both datasets.  
The simple power law turns out to be a good description of the 20--100\,keV spectra in most cases, 
with only 35\% of the sources requiring a more complex fit. The average photon index is around 2 for 
the entire sample and for Seyfert 1 and 2 individually; only Blazars display a slightly flatter spectrum ($\Gamma$$\sim$1.7). 
Interestingly this average spectrum (especially considering Seyfert 1
alone) is steeper than that measured in the CAIXA sample of bright type 1 AGN where $\Gamma$=1.73$\pm$0.09 when 
reflection is included in the fit \cite{bianchi09}; this could be taken as evidence for the presence of a high energy 
cut-off in the  IBIS/BAT waveband. The presence of this (and the reflection) component can be tested via the 
\texttt{pexrav} model applied to those sources where a simple power law is 
not a good fit to the data; although the improvement is significant in most cases, good constraints on all 
spectral parameters are obtained only for NGC 4151 and IC 4329A.
This indicates that, if one wants to get some spectral information from the hard X-ray spectra alone,  
some assumptions must be made on at least one of the 3 parameters involved in the spectral analysis (photon 
index, cut-off energy and reflection). For example in IGR J21247+5058, if we fix the
reflection to zero as found by\cite{molina07, tazaki10}, the photon index and the cut-off
energy can be well constrained ($\Gamma$=1.7$^{+0.1}_{-0.1}$ and E$_{\rm \bf c}$=155$^{+71}_{-37}$\,keV) 
and provide values compatible with previous studies. On the other hand, if in the 
Circinus Galaxy  the high energy cut-off is set to be around 50\,keV as
measured in the past (e.g. \cite{yang09}), the photon index and the reflection can be estimated with some  
precision ($\Gamma$=1.80$^{+0.03}_{-0.03}$ and R=0.59$^{+0.29}_{-0.24}$). 
Finally in the case of 3C 273, where the cut-off is known to be around 1 MeV, the reflection component can be constrained 
to be below 0.3 (see Figure~\ref{2_xx}, right panel).
Unfortunately, such assumptions are not possible in all cases and so one has to rely on broad-band data to obtain
more stringent constraints on source parameters.

\section{Broad-band analysis}
\begin{figure}
\centering
\includegraphics[width=0.37\linewidth]{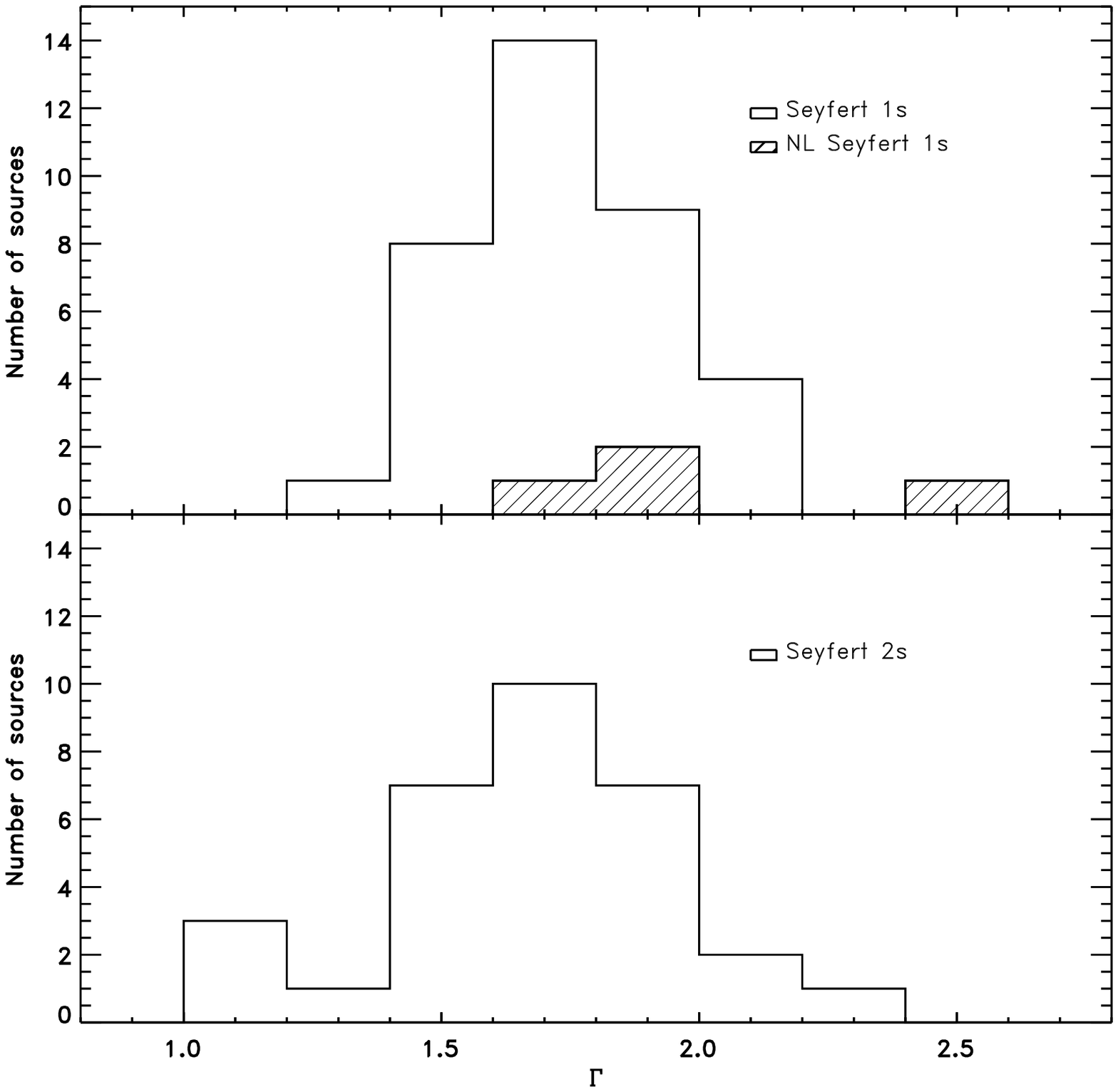}
\includegraphics[width=0.37\linewidth]{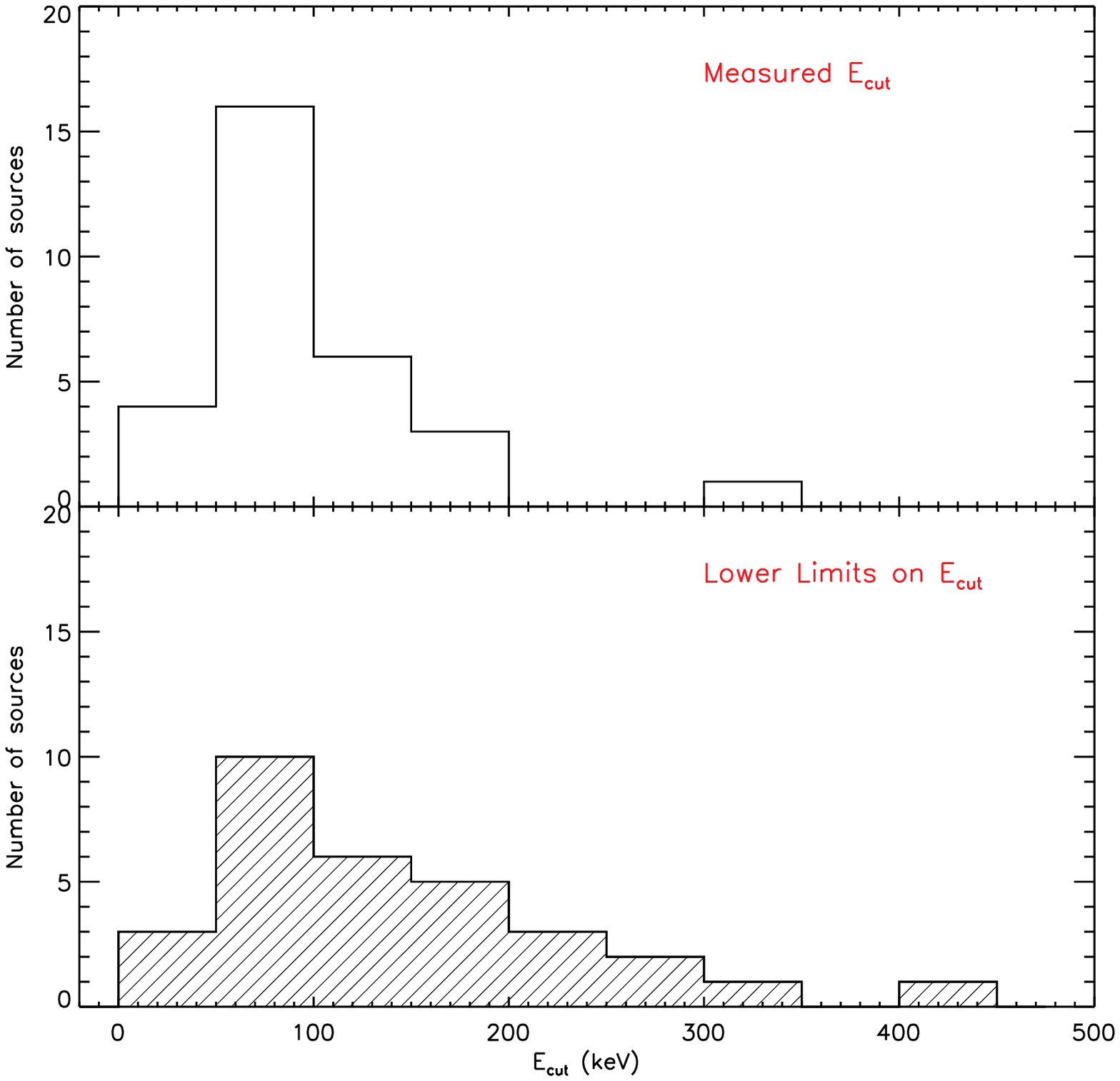}\\
\caption{\small{\emph{Left Panel}: 
Distribution of photon indices of Seyfert 1 (including Narrow Line Seyfert1) at the top and Seyfert 2 at the bottom,
as obtained from broad band X-ray data.
\emph{Right Panel}: Distribution of cut off energies (measured values at the top and lower limits at the bottom)  
for the complete sample of AGN as obtained from broad band X-ray data.} }
\label{gamma_ec}
\end{figure}

A further step towards the comprehension of hard X-ray selected AGN properties is provided by broad-band X-ray
studies. In a series of papers \cite{molina09, panessa08, panessa11, derosa12}, 
we have combined IBIS spectra with data obtained at soft X-ray energies  by various satellites  
such as XMM-Newton, Swift, Chandra etc., to obtain information on various spectral features for most AGN in  the complete sample.
Here, we report a compendium of the results obtained in these papers.
A first result is the great variety in shapes observed and the extreme complexity found when  
modelling most sources; many spectral components have to be considered, although they are not present all the time with
similar strength. The soft excess for example 
is observed in a good fraction of Seyfert 1, but dominates dominates only in few of these AGN; it is 
always present in Seyfert 2, but in 6 objects has a more complex shape than a simple scattered power law.
Absorption is measured in all type 2 objects, while complex absorption is found in 25\% of type 1 sources:
in this last case one or more layers of absorbing material partially covering the source are often found.
The average photon index is typically 1.7, flatter than the canonical value measured in the hard X-ray band but 
compatible with that measured in the CAIXA sample.
Figure \ref{gamma_ec} shows the photon index distribution obtained for type 1 and type 2 
AGN separately: the two distributions as well as the average $\Gamma$ (1.74, $\sigma$=0.2 and 1.68, 
$\sigma$=0.3 for Seyfert 1 and 2 respectively) 
are very similar, which immediately suggests that the production mechanism is the same in 
both types of AGN.

Although high energy spectral components (reflection and 
the high energy cut-off) are not easy to measure due to the  sources spectral complexity and  non simultaneity of the 
soft and hard X-ray data, they have nevertheless being measured in a number of objects.
As is evident in figure \ref{gamma_ec} (right panel),  the distribution of measured cut off energies 
clusters around 100\,keV while the bulk of the lower limits on this parameter are found 
below 300\,keV. One must then conclude that high energy cut-off is present and 
cannot be ignored (as often happens) in broad band fits. This result is in line with a number of 
previous studies \cite{perola02,dadina08,beckmann09}
but is only  barely consistent with those obtained by \cite{ricci11}, who locate the cut-off energy above 
200-300\,keV. 
We also note that our result is  in line with the synthesis models of the cosmic diffuse background (CXB) which
often  assume an upper limit of $\sim$200\,keV for the AGN mean cut-off energy. This choice is basically driven by 
the intensity and  shape of the CXB spectrum above the peak, which cannot be exceeded; even a value of 
300\,keV has difficulties in accommodating all available observations and CXB measurements \cite{gilli07}.

Finally, Figure \ref{gamma_ec} (left and right panels) shows a compendium of the results
obtained for the reflection component in both type 1 and type 2 AGN; note that R is linked to the 
solid angle $\Omega$ as R=$\Omega$/2$\pi$ 
and that measured values and upper limits are shown separately in the figure. 
The distributions of R are similar in both types as well as the average reflection values; it is also evident 
that large values of R are found in highly variable sources and simply reflect the limitation due to the use of non simultaneous data.

\section{Future prospects and conclusions}
 
The number of AGN selected in the hard X-ray band is continuously increasing thanks to both the INTEGRAL and Swift 
surveys and many of the initial expectations have been fulfilled but also  many questions have been raised by the observations made 
so far. Probably the shift in the next few years will be to  multi-waveband studies of these objects which provide 
an alternative view due to their hard X-ray selection: overall IBIS/BAT objects are less affected by absorption than sources 
selected in other wavebands and are representative of the population of high luminosity AGN. Thus  multiwaveband 
studies of them can provide new clues on how AGN work. For example, by combining radio and hard X-ray information 
on a well defined sample of AGN, one can study the jet-disk connection in supermassive 
black holes (Panessa et al. these proceedings) or test if jets are made by e$^{+}$-e$^{-}$ pairs 
\cite{ghisellini12}. Also by surveying hard X-ray selected AGN for water maser emission, one can expect to 
probe the physics of accretion disks, especially in those AGN which are often obscured
at optical/UV and even X-ray wavelengths by large column densities of gas and dust along the line of sight 
\cite{castangia11}. Already the discovery of a water maser feature in the 
nucleus of IGR J16385-2057  has opened a new field of maser studies in Narrow Lines Seyfert 1 galaxies and 
highlighted the possibility of detecting this emission in elliptical galaxies \cite{tarchi11a, tarchi11b}.
As underlined by \cite{elitzur12}, modelling of the infrared spectra of a complete sample of hard X-ray 
selected AGN can provide the only way to measure the intrinsic distribution of torus covering factors and thus 
provide the ultimate test for the clumpy torus model. Finally, one should note that BAT/IBIS samples of AGN show
a large fraction (20-30\%, \cite{koss10}) of systems in interaction, mergers, dual objects etc.
Combined radio/optical/X-ray observations of these systems offer a unique opportunity to study AGN activation 
(see for example \cite{koss12}), probe the relation between interaction and absorption and ultimately 
understand how black holes grow.

This is of course only the beginning of hard X-ray studies of AGN, since INTEGRAL is approved up to 2014 and 
hopefully beyond, Swift/BAT is still in orbit and performing well and Nustar has just joined the group with 
great expectations, but as Plato would say {\textquotedblleft{The beginning is the most important 
part of the work}\textquotedblright}.

\begin{figure}
\centering
\includegraphics[width=0.43\linewidth]{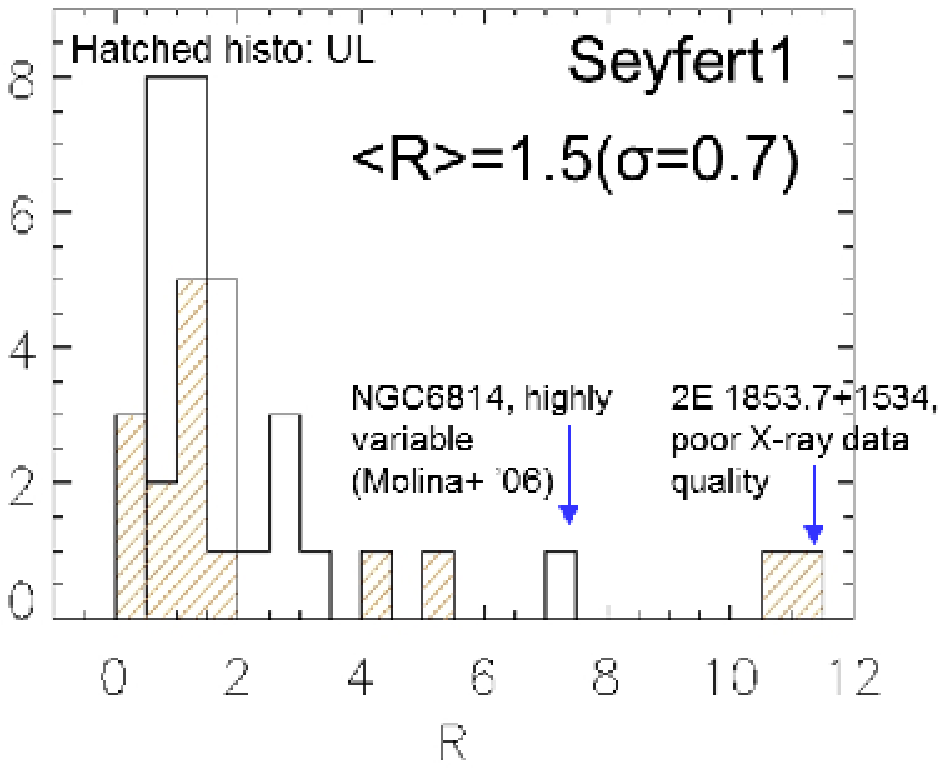}
\includegraphics[width=0.4\linewidth]{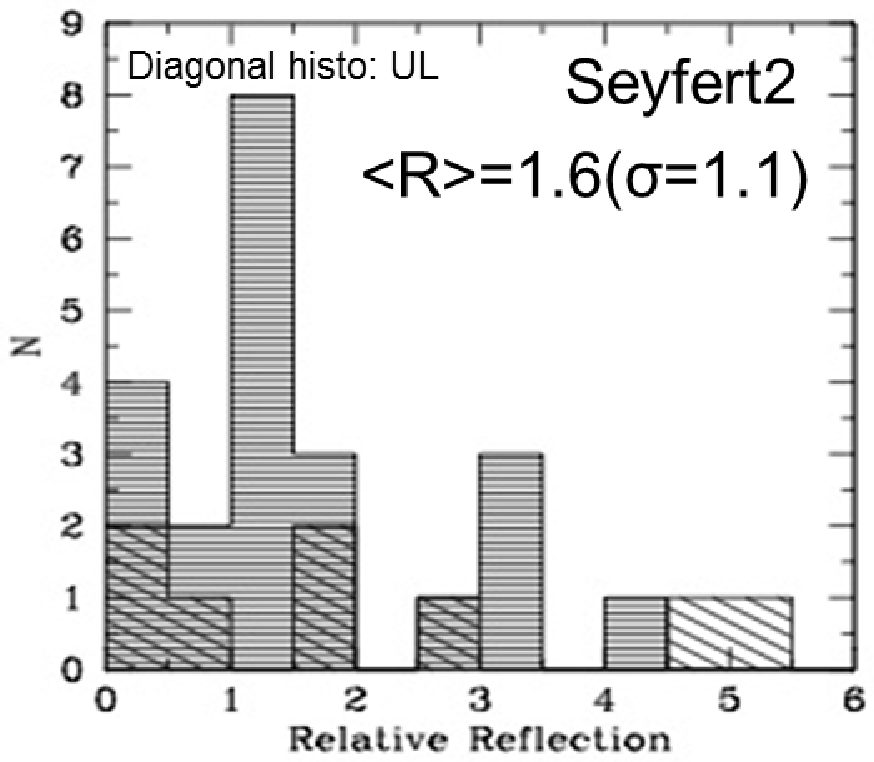}\\
\caption{\small{\emph{Left Panel}: Distribution of the reflection component R
for the Seyfert 1 in the complete sample as obtained from broad band X-ray data; upper limits on R are shown as a diagonal 
histogram. \emph{Right Panel}: Same as in left panel but for Seyfert 2.} }
\label{refl}
\end{figure}

\end{document}